\documentclass[aps,prb,twocolumn,superscriptaddress,groupedaddress]{revtex4-1}
\usepackage{amssymb}
\usepackage{amsmath}
\usepackage{enumitem}
\usepackage[pdftex]{graphicx}
\usepackage{graphicx}
\usepackage{dcolumn}
\usepackage{natbib}
\usepackage{xcolor}

\begin{document}

\title{Colossal Piezoresistance in Narrow-Gap Eu$_5$In$_2$Sb$_6$}

\author{S. Ghosh}
\author{C. Lane}
\author{F. Ronning}
\author{E.D. Bauer}
\author{J.D. Thompson}
\author{J.-X. Zhu}
\author{P.F.S. Rosa}
\author{S.M. Thomas}
\affiliation{Los Alamos National Laboratory, Los Alamos, NM, USA.}

\begin{abstract}
	Piezoresistance, the change of a material's electrical resistance ($R$) in response to an applied mechanical stress ($\sigma$), is the driving principle of electromechanical devices such as strain gauges, accelerometers, and cantilever force sensors.
	Enhanced piezoresistance has been traditionally observed in two classes of uncorrelated materials: nonmagnetic semiconductors and composite structures.
	We report the discovery of a remarkably large piezoresistance in Eu$_5$In$_2$Sb$_6$ single crystals, wherein anisotropic metallic clusters naturally form within a semiconducting matrix due to electronic interactions.
	Eu$_5$In$_2$Sb$_6$ shows a highly anisotropic piezoresistance, and uniaxial pressure along [001] of only 0.4~GPa leads to a resistivity drop of more than 99.95\% that results in a colossal piezoresistance factor of $5000\times10^{-11}$Pa$^{-1}$.
	Our result not only reveals the role of interactions and phase separation in the realization of colossal piezoresistance, but it also highlights a novel route to multi-functional devices with large responses to both pressure and magnetic field.
\end{abstract}


\maketitle

\section{Introduction}

Many semiconducting technologies in use today rely on the control of a material's electrical resistance ($R$) $via$ mechanical strain, electric fields, or magnetic fields.
In particular, the change in resistance as a function of applied mechanical strain, called piezoresistance, is used in a variety of sensing devices~\cite{Barlian2009,Dao2010,Tortonese1993}.
In conventional metals, the piezoresistance effect is typically small and dominated by geometric factors, whereas large piezoresistance can be observed in uncorrelated semiconductors due to changes in electrical resistivity caused by band structure alterations, \textit{e.g.}, carrier mobility, carrier density, and band gap.
Large longitudinal piezoresistive coefficients are observed in silicon, germanium, and silicon carbide~\cite{Smith1954,Kanda1982}.
Interestingly, piezoresistance can be enhanced in nanostructured materials compared to their bulk counterpart.
Silicon nanowires, SmSe thin films, and carbon nanotubes are a few prominent examples of this effect, coined giant piezoresistance effect~\cite{He2006, Copel2013, Maiti2002}. Piezoresistance effects have also been recently observed in phase-separated polymer composites, in which metallic particles, such as carbon nanotubes, are embedded in an insulating matrix.
Large piezoresistance is observed when filler concentrations are close to the percolation threshold, which enables a highly tunable conductive path~\cite{Wang2013}.
Challenges in the control of filler homogeneity and filler-polymer ratio have restricted advances in this field, which invites the question of whether 
alternative phase-separated materials can be designed~\cite{Fiorillo2018}.

Notably, phase separation is a hallmark of strongly correlated materials epitomized in the high-temperature cuprate superconductors and colossal magnetoresistance manganites~\cite{Uehara1999, Emery1993}.
Competing interactions in doped manganites enable the formation of ferromagnetic metallic clusters within a charge-ordered insulating phase~\cite{Jin1994, Elbio2005}.
As a result, the fine balance between electronically distinct phases can be tuned by external parameters, which leads to colossal magnetoresistance (CMR) -- a large reduction in resistance as a function of magnetic fields.
Recent experiments have confirmed early theoretical predictions that dopant-induced disorder is a requirement for the formation of such micrometer-scale phase separation~\cite{Miao2020}.
Mechanical strain has also been shown to be an effective tuning parameter, and large piezoresistance effects have been reported in the manganites~\cite{MohanRadheep2013}.

Motivated by the correspondence between magnetoresistance and piezoresistance in interacting materials, we turn to a special class of CMR materials, namely $f$-electron compounds, in which phase separation occurs even in the absence of dopant-induced disorder.
In EuB$_6$, recent scanning tunneling microscopy experiments directly imaged local inhomogeneities consistent with ferromagnetic clusters~\cite{Pohlit2018}, which stem from the formation of trapped magnetic polarons --- quasiparticles formed at low densities when charge carriers interact strongly with Eu$^{2+}$ spins ~\cite{Penney1972,Mauger1983,Sullow2000,Uehara1999}.
Other $f$-electron examples are $\beta{}$-US$_2$, which shows CMR at low temperature and undergoes an insulator-to-semimetal transition under hydrostatic pressures ~\cite{Ikeda2009}, and $\beta{}$-EuP$_3$, which exhibits one of the largest CMR effects reported to date~\cite{Wang2020}.

\section{Results}
Here we focus on narrow-gap antiferromagnetic semiconductor Eu$_5$In$_2$Sb$_6$, whose recently reported CMR has been attributed to magnetic 
polaron formation \cite{Rosa2020}.
To establish the relationship between piezoresistance and magnetoresistance in Eu$_5$In$_2$Sb$_6$, we perform electrical resistivity measurements under both hydrostatic and uniaxial pressure.
Positive uniaxial pressure is defined to be compressive.
At ambient pressure, Eu$_5$In$_2$Sb$_6$ exhibits semiconducting behavior in electrical resistivity with an apparent activated gap of approximately 40~meV obtained between room temperature and T* $\approx$ 50~K, at which point the activated behavior breaks down.
The room-temperature carrier density was previously found to be $+10^{17}$/cm$^3$.
Prior results, including magnetic susceptibility, magnetoresistance, Hall data, and electron spin resonance argue that the apparent semiconducting behavior is consistent with the formation of magnetic polarons (MP)~\cite{Rosa2020,Souza2022}.

A cartoon depiction of this process is shown in the panels above Fig.~1.
At high temperatures, conduction electrons start to self-trap around Eu$^{2+}$ moments.
This process creates isolated magnetic polarons wherein red regions indicate metallic puddles (region~C).
By depleting the bulk of conduction electrons, the overall resistance displays an activated behavior as long as magnetic polarons are well separated.
At T*, however, inter-polaron interactions set in, and the rise in resistance slows down (region~B).
MP likely have an ellipsoidal shape due to the underlying orthorhombic crystal structure of Eu$_5$In$_2$Sb$_6$.
Future experiments are necessary to determine exactly how the volume-fraction of polarons increases upon cooling.
At $T_{N1}=$~14~K, a percolation path develops, the system orders antiferromagnetically, and the resistance drops markedly (region~A).
As discussed in a previous report~\cite{Rosa2020}, below 9~K the resistance begins to increase again as the temperature is further lowered.
Finally, a second magnetic transition occurs at $T_{N2}=$~7~K.

\begin{figure}[ht]
	\begin{center}
		\includegraphics[width=0.9\columnwidth]{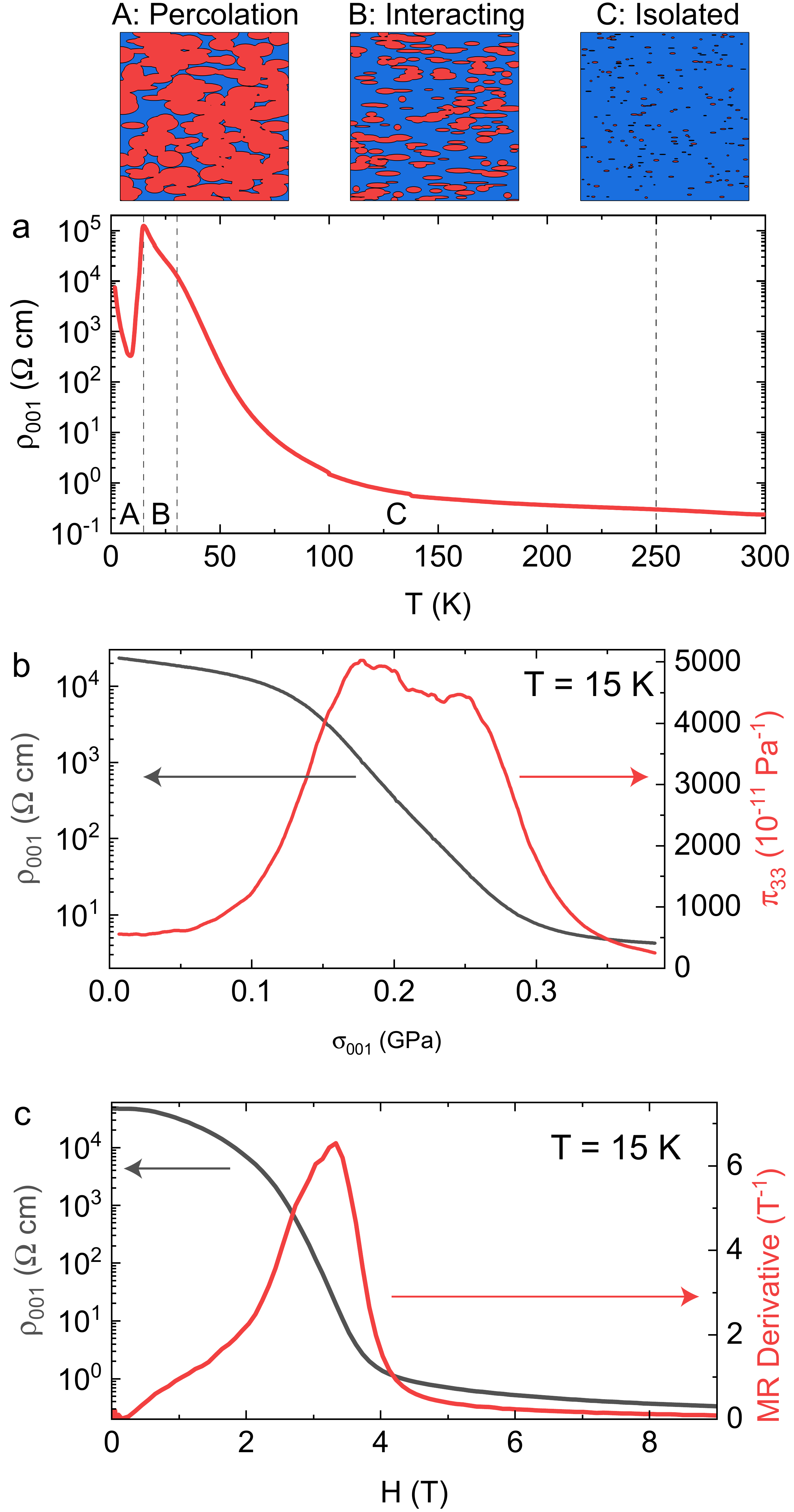}
		\caption{
			Effects of uniaxial pressure on Eu$_5$In$_2$Sb$_6$.
			(a)~Temperature dependent resistivity, $\rho_{001}(T)$, at ambient pressure.
			The cartoons at the top illustrate the evolution of magnetic polarons on cooling, where the labels A, B, and C indicate the relevant temperature ranges.
			(b)~Resistivity and piezoresistance factor versus stress applied along the [001] direction at 15~K.
			\textit{Left axis}:~Resistivity.
			\textit{Right axis}:Stress-dependent piezoresistance coefficient ($\pi_{33}$) as defined in the text.
			Positive uniaxial stress indicates compression.
			(c)~Resistivity and MR derivative, $-\frac{dR}{dH}/R\left(H\right)$, as a function of field applied along [010] at 15~K.
			\textit{Left axis}:~Resistivity.
			\textit{Right axis}:~MR derivative.
		}
		 \label{fig:fig1}
	\end{center}
\end{figure}

Our results show that Eu$_5$In$_2$Sb$_6$ exhibits both colossal piezoresistance and CMR effects just above T$_{N1}$.
Figure~1(b) shows both the resistivity and the stress-dependent piezoresistance coefficient, $\pi=-\frac{d\rho}{d\sigma}/\rho\left(\sigma\right)$, as a function of [001] stress at 15~K.
Note that the piezoresistance is normalized by the resistivity as a function of stress instead of at zero stress because of the large decrease in the resistivity with applied stress.
To measure $\pi_{33}$, stress is applied along [001] and resistivity is measured along [001].
In comparison, Fig.~1(c) shows the magnetoresistance at 15~K and a similar peaked shape in the MR analogue of the piezoresistance coefficient: $-\frac{dR}{dH}/R\left(H\right)$.
The similarities between colossal piezoresistance and MR in the same temperature window indicate that both strain and magnetic field are effective in driving magnetic polarons through a percolation threshold.

Under [001] stress, the resistance decreases by 99.95\% over a change in stress of just under 0.4~GPa.
The peaked shape of $\pi_{33}$ shows that the formation of a percolation path of MP is confined to a narrow range of stresses between 0.1 and 0.35~GPa.
Notably, the piezoresistance coefficient in Eu$_5$In$_2$Sb$_6$ saturates near a value of $556$~$10^{-11}$Pa$^{-1}$ for uniaxial stress less than 0.05~GPa.
This may reflect the intrinsic piezoresistance of the narrow-gap semiconducting state and is still a factor of two larger than in silicon~\cite{Smith1954}.
At stresses larger than 0.35~GPa, the magnitude of the $\pi$-coefficient continues to decrease slowly as stress is increased.
This is because a percolation path has been fully formed by stress and most of the current flows through a low-resistance region in the sample.
As shown in Fig.~2(a), the uniaxial stress range where a peaked shaped is observed in the piezoresistance coefficient becomes less narrow as the system moves away from the percolation threshold temperature, but the effect persists to at least 25~K.
In addition, Fig.~2(b) shows that the percolation path formation is also fully reversible as the uniaxial pressure is tuned at fixed temperature.

\begin{figure*}[ht]
	\begin{center}
		\includegraphics[width=1\textwidth]{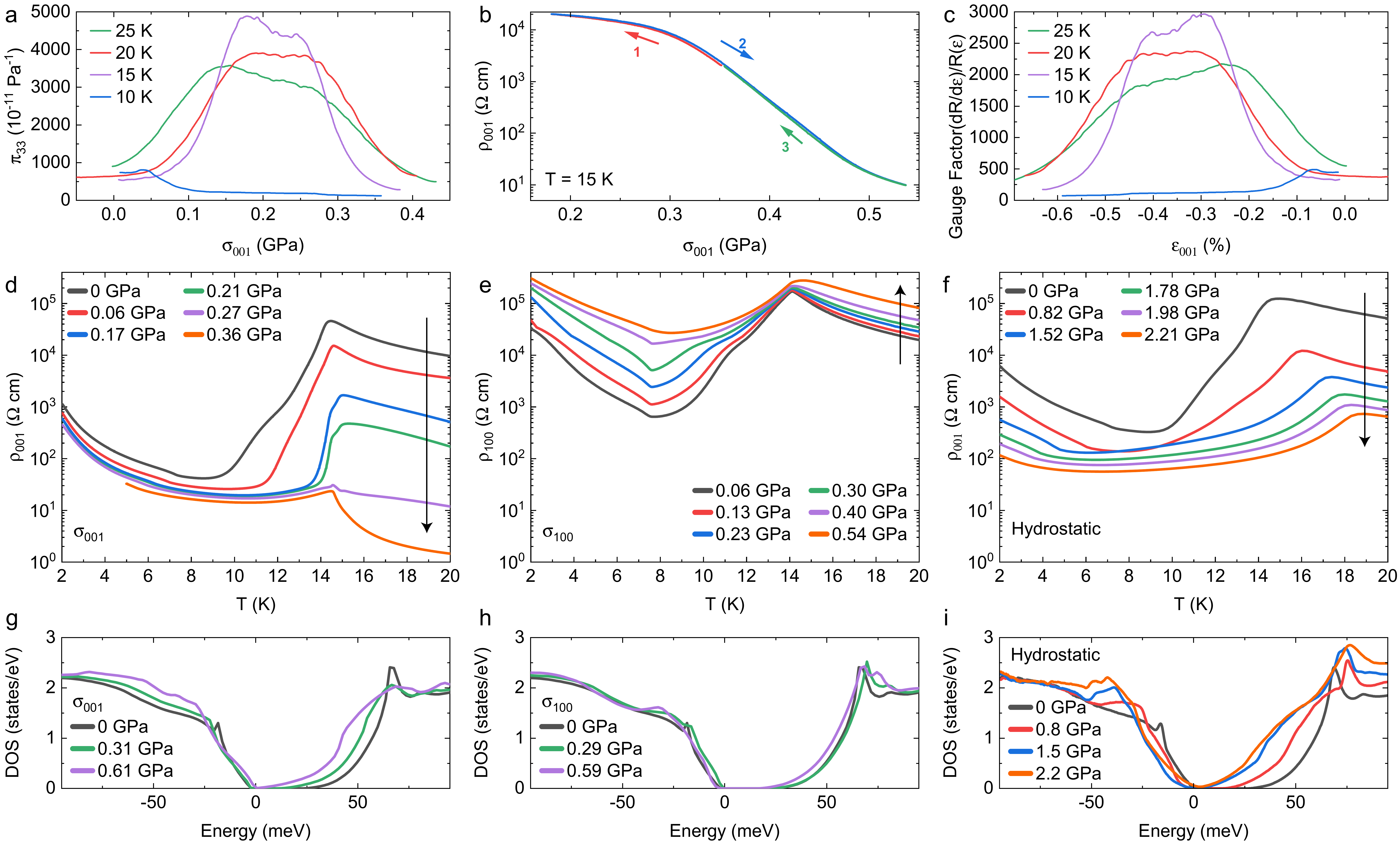}
		\caption{
			Top row: uniaxial pressure dependence of Eu$_5$In$_2$Sb$_6$ at fixed temperatures.
			(a)~Stress-dependent piezoresistance coefficient ($\pi_{33}$).
			(b)~Reversibility of colossal piezoresistance.
			(c)~Estimated gauge factor (see text). Negative strain is compression.
			Middle row: low temperature resistivity versus temperature of Eu$_5$In$_2$Sb$_6$ under various pressure conditions.
			Subscript of $\rho$ indicates direction of applied current.
			The arrows indicate the direction of increasing pressure.
			(d)~Uniaxial pressure along [001] ($\sigma_{001}$), (e) uniaxial pressure along [100] ($\sigma_{100}$), and (f)~hydrostatic pressure.
			Bottom row: calculated density of states versus energy for various pressure conditions.
			(g)~Uniaxial pressure along [001] ($\sigma_{001}$), (h) uniaxial pressure along [100] ($\sigma_{100}$), and (i)~hydrostatic pressure.
		}
		\label{fig:Fig2}
	\end{center}
\end{figure*}

To estimate the gauge factor, we use the room-temperature value of the Young's modulus ($E=60.7$~GPa) reported for polycrystalline Yb$_5$In$_2$Sb$_6$.
This non-magnetic analogue can give a reasonable estimate to convert between stress and strain.
Figure~2(c) shows the estimated strain-dependent gauge factor for several different temperatures.
At 15~K, the estimated gauge factor in bulk Eu$_5$In$_2$Sb$_6$ reaches approximately 3000, which is the record value for bulk materials and comparable to the highest reported value in silicon nanowires of approximately 5000~\cite{He2006, Barwicz2010}.
It should be noted that more recently, the gauge factor in silicon nanowires was shown to be much smaller and closer to the bulk value, with the large value in initial reports being attributed to experimental artifacts~\cite{Milne2010, Lugstein2010}.

Interestingly, Eu$_5$In$_2$Sb$_6$ displays highly anisotropic and temperature-dependent behavior under uniaxial pressure.
Figures~2(d) and (e) show the electrical resistivity as a function of pressure applied along the [001] and [100] directions, respectively.
The highest sensitivity is observed for uniaxial pressure applied along [001], wherein the largest piezoresistance effect is confined to temperatures above $T_{N1}$. 
At 20~K, the resistivity is reduced by nearly four orders of magnitude with only 0.36~GPa of applied uniaxial pressure.
This is naturally explained by the pressure-induced percolation of anisotropic metallic clusters (MP).

Additionally, a small splitting of the magnetic transition at $T_{N1}$ is observed only under uniaxial pressure along [001], and this effect is most evident at 0.17 and 0.27~GPa.
There are two possible scenarios for this splitting.
First, another magnetically ordered state may appear under uniaxial pressure.
Second, uniaxial pressure may split the percolation threshold temperature and T$_{N1}$ so that percolation occurs at a slightly higher temperature than T$_{N1}$, similar to the case observed in EuB$_6$~\cite{Sullow2000}.
For all directions, there is also an inflection in the zero pressure data near 12~K.
From prior heat capacity and magnetization measurements, this feature near 12~K is not related to any thermodynamic transition~\cite{Rosa2020}. 

\begin{figure}[ht]
    \centering
    \includegraphics[width=1\columnwidth]{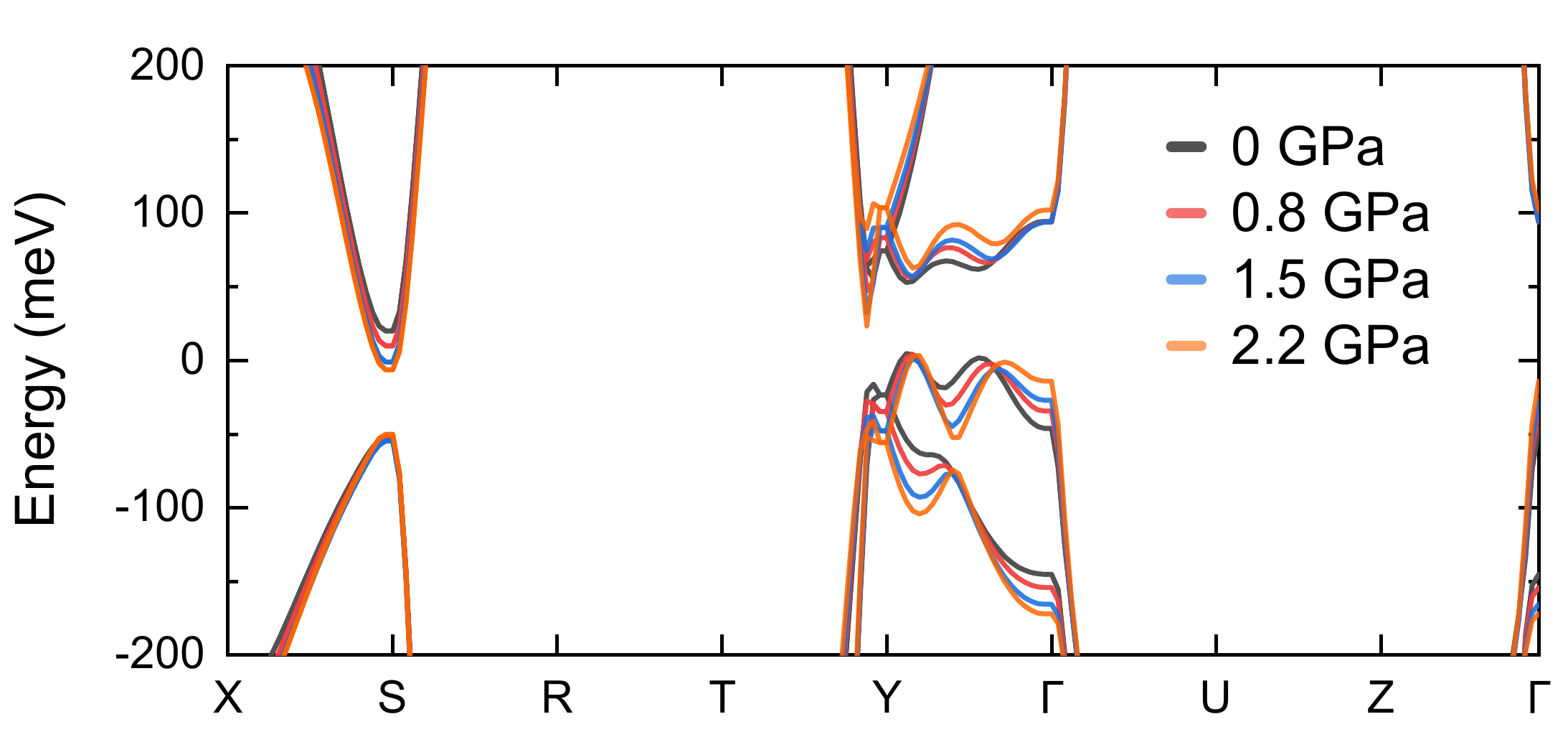}
    \caption{\label{fig:Fig3}
        Hydrostatic pressure effects on band structure.
    }
\end{figure}

For uniaxial pressure along the [100], the pressure dependence of the resistance is opposite to the [001] case above $T_{N1}$, $i.e.$, the resistance \textit{increases} by a factor of 4 as the pressure is increased from 0.06~GPa to 0.54~GPa.
Near $T_{N2}=$~7~K, however, a temperature at which Eu$_5$In$_2$Sb$_6$ enters a different antiferromagnetic state, the uniaxial pressure dependence along [100] is greatly enhanced.
The same change in pressure from 0.06~GPa to 0.54~GPa increases the resistivity by nearly two orders of magnitude.
As the temperature is further lowered, this effect once again becomes less pronounced.

In contrast, Eu$_5$In$_2$Sb$_6$ exhibits a large reduction in resistivity across the entire temperature range under hydrostatic pressure.
Figure~2(f) highlights its low-temperature behavior.
A two order-of-magnitude reduction in resistivity is observed with an applied pressure of 2.2~GPa across nearly the entire measured temperature range, with a slight deviation near the minimum resistivity at 10~K.
This effect is qualitatively different from uniaxial pressure along either [001] or [100], in which the largest changes in resistivity occurred either above or below $T_{N1}$, respectively, and it is consistent with the presence of anisotropic metallic clusters.

Notably, changes to the band structure of Eu$_5$In$_2$Sb$_6$ due to hydrostatic or uniaxial pressure cannot account for our results.
We performed electronic structure calculations using density functional theory (DFT) to determine the density of states as a function of hydrostatic and uniaxial pressure in the paramagnetic state.
As shown in the bottom row of Fig.~2, for all pressure configurations the size of the DFT gap decreases as pressure is increased.
In addition, Fig.~3 shows how the band structure changes as function of hydrostatic pressure, where the most notable change is the hole pocket at the S point moving towards the Fermi energy.
These results are inconsistent with data for uniaxial pressure applied along [100], wherein the resistance increases with applied pressure.
Although changes in mobility may counteract a decrease in the size of the gap, a more natural explanation for the resistance increase is a stress-induced change in the current path between MP.
This will be discussed further below.

\begin{figure}[ht]
    \centering
    \includegraphics[width=1\columnwidth]{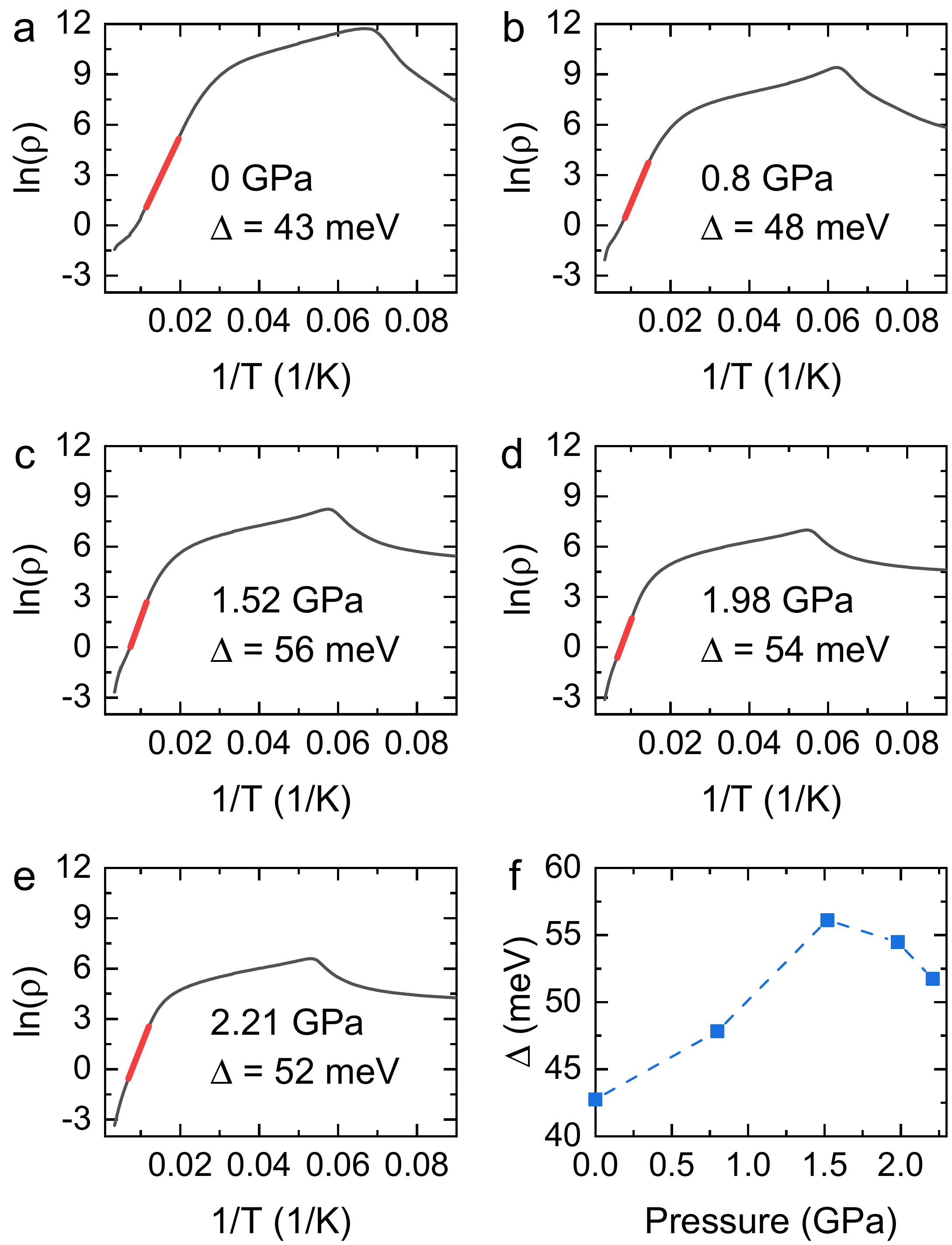}
    \caption{\label{fig:Fig4}
	(a)--(e)~Arrhenius fits at a number of different hydrostatic pressures.
	As noted in the test, the fit range is quite narrow in temperature.
	(f)~The apparent gap as a function of pressure.
    }
\end{figure}

The gap extracted from Arrhenius fits to the intermediate-temperature region of the resistivity under hydrostatic pressure shows that the apparent gap surprisingly increases slightly as a function of pressure, in contrast to the DFT results.

These fits are shown in Fig.~4.
The fit range is quite narrow, with the higher temperature limit ranging from about 200~K to 100~K and the lower temperature limit ranging from 100~K to 50~K. At low pressure and at high temperatures there is a slope change near 200~K, which corresponds to the beginning of MP formation. At low temperatures, there is a broad slope change that corresponds to the temperature where MP begin to interact. This occurs near 50~K. Both of these features move to higher temperature with increasing pressure. In all cases, the fit only covers around half a decade in temperature.
Care should therefore be taken when interpreting these fits as they only take place over a narrow temperature range, but they do show that there is no clear trend of the size of the gap decreasing under hydrostatic pressure.

DFT calculations also cannot fully explain the extremely large piezoresistance observed for uniaxial pressure applied along [001].
The gap reduction in DFT is monotonic as a function of uniaxial pressure.
Thus, the peaked-shape in the piezoresistance coefficient cannot be explained by changes in the density of states.
Further, similarly to hydrostatic pressure, the temperature dependence of the resistivity does not show any indication that the gap closes and the system becomes metallic.
Nonetheless, DFT calculations do show some consistency with the pressure results.
The effect of pressure on the DOS is largest for uniaxial pressure along [001], followed by hydrostatic pressure and [100] pressure.
This hierarchy is not well-reflected in the MP percolation temperature, however.
In principle, a larger DOS should increase the number of electrons available to self-trap around Eu moments, leading to larger polaron size at a given temperature.
This should then increase the temperature at which percolation occurs.
For [001] pressure, the percolation temperature remains near 15~K at all pressures measured, even though the resistance in the paramagnetic state decreases by several orders of magnitude. 

Finally, we note that the resistivity of Eu$_5$In$_2$Sb$_6$ at room temperature does decrease under hydrostatic pressure, leading to a lower overall resistance as would be expected from a reduction in the band gap.
Because DFT does not take into account magnetic polaron or inhomogeneity effects, our results therefore indicate that near room temperature --- where magnetic polarons do not affect electronic transport --- DFT may be able to successfully predict a decrease in the activated gap with pressure.

\begin{figure}
	\begin{center}
		\includegraphics[width=.95\columnwidth]{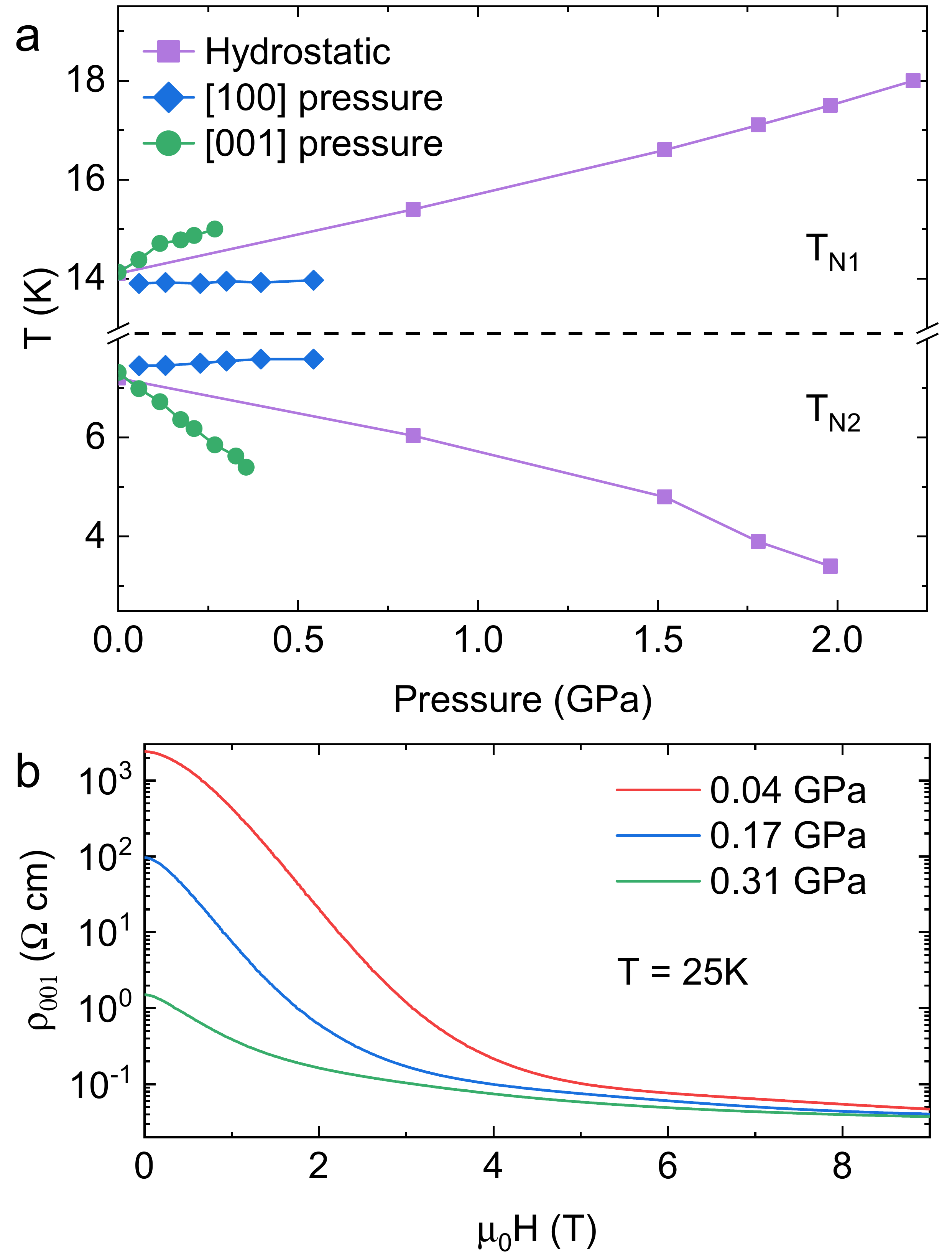}
		\caption{
			(a)~Pressure-temperature phase diagram for hydrostatic and uniaxial pressures.
			(b)~Magnetoresistance with different amounts of uniaxial pressure applied along [001] at 25~K and field applied along [010].
		}
		\label{fig:fig5}
	\end{center}
\end{figure}

We also rule out changes in the overall pressure-temperature phase diagram as the source of the colossal piezoresistance.
Figure~5(a) shows the phase diagram of Eu$_5$In$_2$Sb$_6$ for both hydrostatic and uniaxial pressure.
Under hydrostatic pressure, T$_{N1}$ increases at a rate of +1.8(1)~GPa/K, whereas T$_{N2}$ decreases at a rate of -1.9(2)~K/GPa.
As will be shown later, this pressure dependence matches very well the expected pressure dependence determined from thermal expansion.
For hydrostatic pressure, the largest changes in resistance at a given temperature versus pressure are centered near T$_{N1}$. 
The resistance change is enhanced by the fact that T$_{N1}$ moves to higher temperature as pressure is increased.
In comparison, for the case of [001] pressure, the largest piezoresistance occurs in a range of temperatures above T$_{N1}$.
This is further evidenced by the fact that an overall change in resistance of 99.99\% still occurs at 20~K, as shown by the large piezoresistance coefficient at 20~K in Fig.~2(a), and is therefore not directly caused by the onset of magnetic order.
As temperature is increased above T$_N$, however, the percolation occurs over an increasingly wider range of uniaxial [001] pressure.

Our combined results can be consistently explained by the percolation of anisotropic magnetic polarons.
This anisotropy is likely caused by a combination of the orthorhombic crystal structure and anisotropic magnetic interactions.
Similar behavior was observed in a thin-film manganite grown on a mismatched substrate that resulted in anisotropic strain~\cite{Ward2009}.
In that case, the anisotropic strain resulted in both a resistance anisotropy and a difference in the metal-to-insulator transition temperature depending on the direction of applied current.
Here, we do not see a different transition temperature between the magnetic order and the metallization transition, even though we do observe vastly different behavior between the [100] and [001] directions.

To further confirm the percolation scenario, we measured magnetoresistance with field applied along [010] at a number of fixed uniaxial pressures applied along [001].
As shown in Fig.~5(b), as the stress is increased, the CMR is rapidly suppressed.
This is precisely the result expected from uniaxial pressure driving magnetic polarons closer to percolation.
Although magnetic field is expected to increase the size of MP due to spin polarization of the ferromagnetic clusters, its effects will be reduced when a percolation path is already established via pressure.
Similarly, above 6~T the piezoresistance is greatly reduced compared to zero field, which also reveals that magnetic field has driven the MP through percolation.
In effect, both applied stress along the appropriate direction and magnetic field can be used to tune the system through percolation.
Directly imaging magnetic polarons in Eu$_5$In$_2$Sb$_6$, however, is challenging.
Though its highly insulating behavior hinders scanning tunneling microscopy measurements{~\cite{Crivillero2022}, small-angle neutron scattering measurements could be valuable in determining the size of the magnetic polarons. 

\begin{figure}
    \centering
    \includegraphics[width=1\columnwidth]{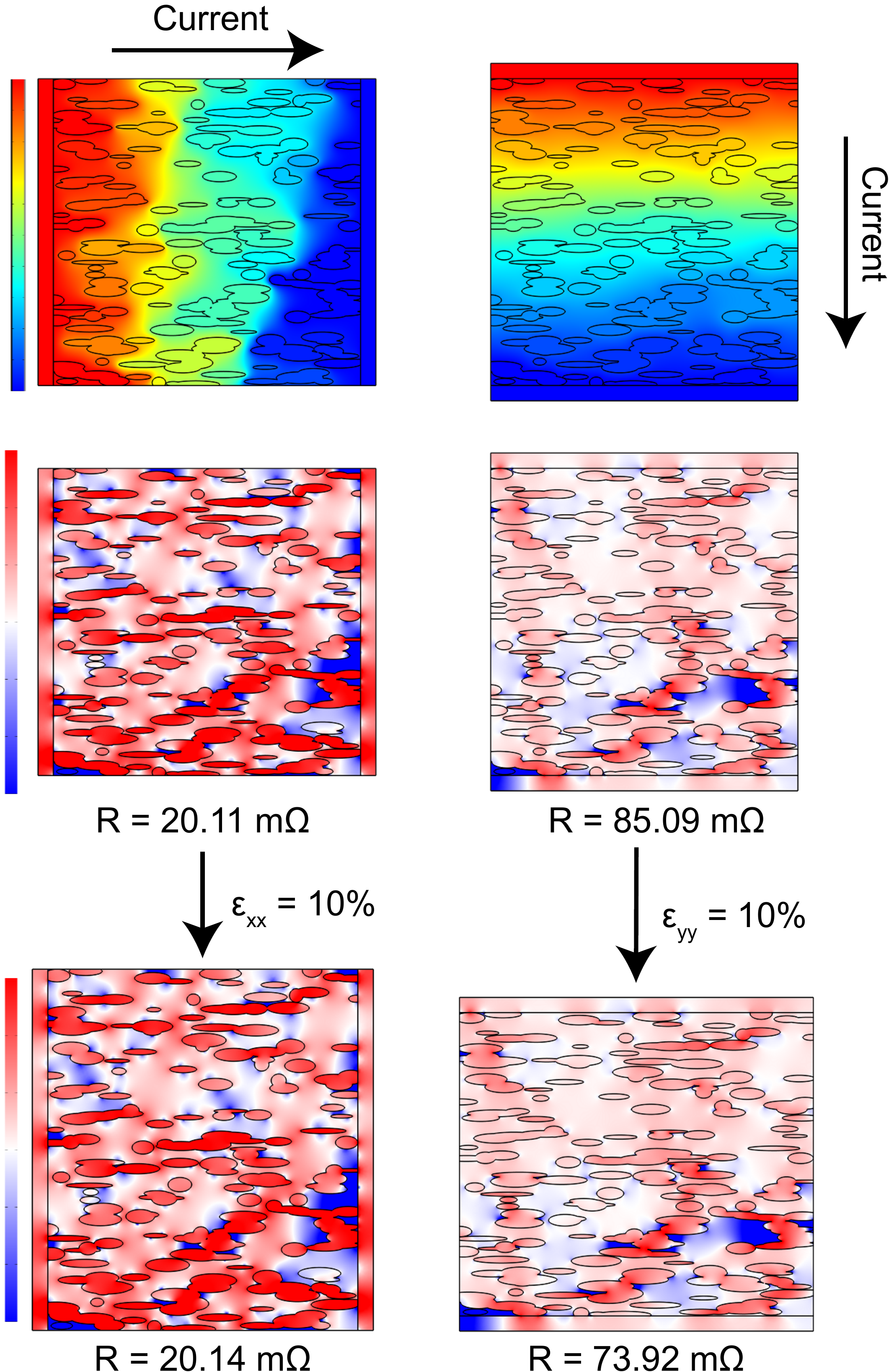}
    \caption{
        \label{fig:sfig4}
        Simulation of effects of strain on resistance in a mixed-phase material with anisotropic shape for MP.
        For the left column, the color represents electric potential on a linear scale (color bar on left).
        For the remaining plots, the color represents current density on a log scale (color bar on right).
		See text for more details.
    }
\end{figure}

For magnetic polarons with an ellipsoidal shape, a simple model suggests the resistivity will be most sensitive to stress applied along the shortest axes of the ellipsoid and may increase for stress applied along the longer axes.
We performed finite element simulations of a two-dimensional mixed-phased material, as shown in Fig.~6.
The ellipses represent the metallic MP phase and are randomly generated with longer dimension along $x$ and a shorter dimension along $y$.
The insulating background has resistivity seven orders of magnitude larger than the MP phase.

Current flows either along the $x$ or $y$ direction.
The top row shows the calculated electric potential for these two conditions.
The middle row shows the current density on a logarithmic scale and the calculated resistance.
In the bottom row, the sample is strained by 10\% in either the $x$ or $y$ direction, and the current density is calculated for the deformed geometry.
The strain direction is parallel to the direction of applied current.
Once again, the resistance is determined.

For current and strain parallel to the $x$ direction the resistance increases slightly upon applying strain.
This can be understood intuitively if one considers the effects of Poisson's ratio.
A compression in the $x$ direction results in an expansion in the $y$ direction.
Due to the anisotropic MP shape, this serves to increase the current path length between MP, which is mainly in the $y$ direction.
This leads to a slight increase in overall resistance.
Conversely, a compression in the $y$ direction directly decreases the path between MP, leading to a decrease in resistance.
This simulation is limited because it does not consider the effects of strain on the size and shape of the MP, nor does it consider piezoresistance of the insulating background.
Nonetheless, it provides a simple example of how MP with anisotropic shape can account for the observed strain dependence in E$_5$In$_2$Sb$_6$.

We thus expect the MP to have an anisotropic shape with the shortest axis along [001] and the longest axis along either [100] or [010], which is consistent with a magnetic structure with no spin component along [001].
Neutron and x-ray diffraction measurements, which confirm that the Eu spins lie in the (001) plane, will be reported elsewhere. 

\begin{figure}
	\begin{center}
		\includegraphics[width=.95\columnwidth]{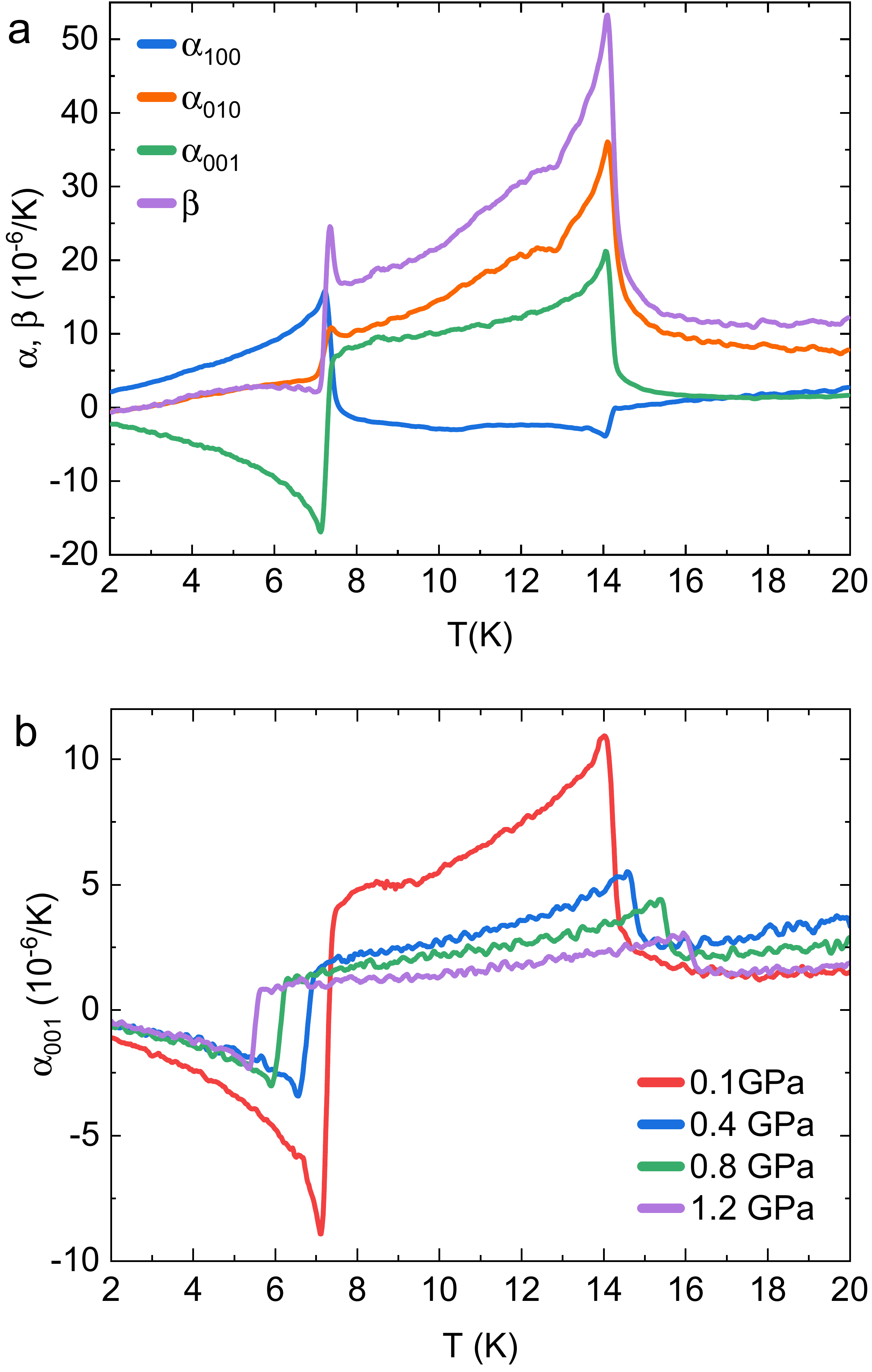}
		\caption{
			(a)~Low-temperature thermal expansion of Eu$_5$In$_2$Sb$_6$ at ambient pressure.
        	(b)~Low-temperature thermal expansion of Eu$_5$In$_2$Sb$_6$ along the $c$ axis at different hydrostatic pressures.
		}
		\label{fig:fig7}
	\end{center}
\end{figure}

To better understand the pressure-temperature phase diagram, we performed thermal expansion measurements, which are shown in Fig.~7.
The Ehrenfest relation can be used to determine the pressure dependence of a second order phase transition:
\begin{equation}
	\frac{\partial{T}}{\partial{P}}=\frac{\Delta{}\beta {}V_{m}}{\Delta{}C/T}
\end{equation}
\noindent Here, $\Delta\beta$ is the jump in volume thermal expansion coefficient, $\Delta{}C/T$ is the jump in heat capacity divided by temperature, and $V_m$ is the molar volume.

The value for $\partial{T}/\partial{P}$ can be calculated by combining the thermal expansion data in Fig.~7(a) with the heat capacity data from Rosa \textit{et al.}\cite{Rosa2020}. 
This gives a value of +2.0(1)~K/GPa for $T_{N1}$ and -2.1(1)~K/GPa for $T_{N2}$, in reasonable agreement with the pressure-temperature phase diagram for hydrostatic pressure shown in Fig.~5(a) and reported above.

For the uniaxial pressure dependencies, care must be taken when attempting to use a one-dimensional version of the Ehrenfest relations.
For example, at T$_{N2}$ the jump in $\Delta\alpha_{100}$ is approximately $+17\times 10^{-6}K^{-1}$, whereas the jump is $\Delta\alpha_{001}$ is approximately $-26\times 10^{-6}K^{-1}$.
This suggests that T$_{N2}$ should increase for $a$-axis pressure and decrease with $c$-axis pressure but with a slightly higher rate.
Instead, T$_{N2}$ increases at only $+0.3(1)$~K/GPa for $a$-axis pressure, but decreases very quickly at $-5.3(1)$~K/GPa for $c$-axis pressure.
At T$_{N1}$, the transition temperature shows effectively no change with $a$-axis pressure ($+0.1(1)$~GPa/K) and increases at $+3.2(4)$~K/GPa with $c$-axis pressure.
Note that T$_{N1}$ could only be tracked up to 0.27~GPa for $c$-axis pressure due to the change in the behavior of resistivity versus temperature at higher pressures.

\section{Conclusions}

We have shown that Eu$_5$In$_2$Sb$_6$ exhibits highly anisotropic colossal piezoresistance that is driven by the percolation of magnetic polarons.
The presence of both colossal magnetoresistance and colossal piezoresistance in Eu$_5$In$_2$Sb$_6$ can be exploited in future multi-functional devices, and it highlights the promise of interacting systems in the search for novel colossal piezoresistance materials.
Because these effects are occurring in a clean, stoichiometric compound, Eu$_5$In$_2$Sb$_6$ may serve as a keystone to understanding colossal magnetoresistance and piezoresistance in a broad class of materials wherein similar mixed-phase effects emerge.

\section{Methods}

Single crystals of Eu$_5$In$_2$Sb$_6$ were grown by In-Sb self-flux method \cite{Rosa2020}.
The crystal structure of the Zintl antimonide Eu$_5$In$_2$Sb$_6$ shares the orthorhombic Ca$_5$Ga$_2$As$_6$-type with space group P\textit{bam}.
Here every In atom is coordinated by four Sb atoms, forming a InSb$_4$ tetrahedron.
These tetrahedra are then linked along the corner through the Sb–Sb bonds~\cite{Park2002}.

Measurements under uniaxial pressure were performed using a commercially available uniaxial pressure cell (Razorbill FC-100) with samples mounted with both pressure and current leads along the same axis.
Sample dimensions allowed for measurements only along the [100] and [001] directions.
The sample dimensions along [010] were too small to perform similar measurements with stress applied along [010].
For each of the measurements, current was applied in the same direction as the stress.
To calculate the gauge factor, uniaxial pressure was converted to strain using reported room-temperature values of the elastic constants ($E=60.7$~GPa) for the non-magnetic analogue Yb$_5$In$_2$Sb$_6$~\cite{Aydemir2015}.
Hydrostatic pressure measurements were performed using a piston clamp pressure cell.
Resistance measurements were performed using a standard 4-point technique, whereas thermal expansion measurements under pressure were performed using the technique described in Rosa \textit{et al.}~\cite{Rosa2017}.
Ambient pressure thermal expansion measurements were performed using a capacitance cell dilatometer~\cite{Schmiedeshoff2006a}.

Density functional theory based first-principles electronic structure calculations were carried out by using the pseudopotential projector-augmented wave method~\cite{Kresse1999} implemented in the Vienna ab initio simulation package (VASP)~\cite{Kresse1996,Kresse1993}.
We used an energy cutoff of $500$ eV for the plane-wave basis set.
Exchange-correlation effects were treated using the Perdew-Burke-Ernzerhof (PBE) GGA density functional~\cite{Perdew1996}, where a 9 $\times$8$\times$25~$\Gamma$-centered k-point mesh was used to sample the Brillouin zone.
Spin-orbit coupling effects were included self-consistently.
A $Pbam$ (Space group number: 55) crystal structure in accordance with the experimental measurements~\cite{Park2002} was used throughout the calculations.
For each hydrostatic (uniaxial) pressure, all atomic sites in the unit cell along with the unit cell dimensions were relaxed simultaneously using a conjugate gradient algorithm to minimize energy with an atomic force tolerance of $0.001$ eV/\AA~and a total energy tolerance of $10^{-8}$ eV.

\begin{acknowledgments}
Experimental work at Los Alamos was supported by U.S. Department of Energy, Office of Basic Energy Sciences, Division of Materials Science and Engineering project “Quantum Fluctuations in Narrow-Band Systems.”
S.~Ghosh, C.~Lane, and J.-X.~Zhu were supported by the Los Alamos Laboratory Directed Research and Development program.
Scanning electron microscopy, focused ion beam milling, and femtosecond laser machining were performed at the
Center for Integrated Nanotechnologies, an Office of Science User Facility operated for the U.S. Department of Energy Office of Science.
\end{acknowledgments}

\bibliography{lib.bib}

\end{document}